\documentclass[twocolumn]{aastex62}
\usepackage{graphics,epsf}
\usepackage{amsmath}                
\usepackage{amsfonts}               
\usepackage{amssymb}                
\usepackage{epsfig}                 
\usepackage{appendix}
\usepackage{graphicx}
\usepackage{float}
\usepackage{color}
\usepackage{multirow}
\usepackage[para,online,flushleft]{threeparttable}

\newcommand{\km}{{~\rm km}}
\newcommand{\s}{{~\rm s}}

\newcommand{\K}{{~\rm K}}

\newcommand{\yr}{{~\rm yr}}

\newcommand{\pc}{{~\rm pc}}

\newcommand{\days}{{~\rm days}}

\begin{document}

\title{On rare core collapse supernovae inside planetary nebulae}

\email{ealealbh@gmail.com; soker@physics.technion.ac.il}

\author{Ealeal Bear}
\affiliation{Department of Physics, Technion – Israel Institute of Technology, Haifa 3200003, Israel}

\author{Noam Soker}
\affiliation{Department of Physics, Technion – Israel Institute of Technology, Haifa 3200003, Israel}
\affiliation{Guangdong Technion Israel Institute of Technology, Guangdong Province, Shantou 515069, China}

\begin{abstract}
We conduct simulations using \textsc{MESA} of the reverse formation of a white dwarf (WD)-neutron star (NS) binary system in which the WD forms before the NS. We conclude that a core collapse supernova (CCSN) explosion might occur inside a planetary nebula (PN) only if a third star forms the PN.
In this WD-NS reverse binary evolution, the primary star evolves and transfers mass to the secondary star, forms a PN, and leaves a WD remnant. If the mass-transfer brings the secondary star to have a mass of $ \ga 8 M_\odot$ before it develops a helium core, and if the secondary does not suffer an enhanced mass-loss before it develops a massive helium core, e.g., by mass-transfer, it explodes as a CCSN and leaves a NS remnant.
The time period from the formation of the PN by the primary to the explosion of the secondary is $\ga 10^6 \yr$. By that time the PN has long dispersed into the interstellar medium.
In a binary system with nearly equal mass components,
the first mass-transfer episode takes place after the secondary star has developed a helium core and it ends its life forming a PN and a WD. 
The formation of a CCSN inside a PN (CCSNIP) requires the presence of a third star. The third star should be less massive than the secondary star but by no more than ${\rm few} \times 0.01 M_\odot$. We estimate that the rate of CCSNIP is $\approx 10^{-4}$ times the rate of all CCSNe.
\end{abstract}

\keywords{planetary nebulae: general – stars: massive - binaries (including multiple): close -  supernovae: general} 

\section{Introduction} 
\label{sec:intro}

The idea that some type Ia supernovae (SNe Ia) explode inside a planetary nebula (PN; \citealt{DickelJones1985}), so called a SN inside a PN (SNIP; e.g., \citealt{TsebrenkoSoker2013, TsebrenkoSoker2015}), is natural since SNe Ia are exploding white dwarfs (WDs). In the present study we ask whether a core collapse supernova (CCSN) might take place inside a PN, and if yes, then what are the conditions for that to occur. Such a scenario requires the WD to be born before the neutron star (NS), and therefore this binary evolutionary channel is termed a {\it WD-NS reverse evolution}. The initial more massive star in the binary system, the primary star, has a mass of $M_{\rm 1,B, i} \la 8-9 M_\odot$. It evolves first, transfers mass to its companion, and forms the WD. The mass-transfer brings the secondary to have a post-accretion mass of $M_{\rm 2,B, f} \ga 8 M_\odot$, and so it might end its life as a CCSN, leaving behind a NS remnant. 

In this paper we will refer to the initial more massive star as the `primary star', and to the initial less massive star as the `secondary star'. This definition does not depend on which star loses mass and which star gains mass, and we keep this definition even when the primary star becomes less massive than the secondary star.  

Many earlier studies considered 
evolution that leaves behind a WD-NS binary system, or a single NS, where a WD is formed before the NS does, e.g., a WD experiences an accretion induced collapse (AIC) to form a NS or some similar processes (e.g. \citealt{Ruiteretal2019, TaurisJanka2019, LiuWang2020, WangLiu2020}).
Some studies consider a mass-transfer process that brings the secondary star to evolve toward a CCSN, leaving behind a WD-NS system, bound or unbound 
(e.g.,  \citealt{TutukovYungelSon1993, PortegiesZwartVerbunt1996,  TaurisSennels2000, Brownetal2001, Nelemansetal2001, Kimetal2003, Kalogeraetal2005, vanHaaftenetal2013, Toonenetal2018, Breiviketal2020}). The WD-NS reverse evolution might account for a massive WD in the binary radio pulsars PSR~B2303+46 and PSR~J1141-6545 (e.g., \citealt{PortegiesZwartYungelson1999, vanKerkwijkKulkarni1999,  TaurisSennels2000, Brownetal2001, Daviesetal2002, Churchetal2006}). 
\cite{SabachSoker2014} mentioned other outcomes of the WD-NS reverse evolution, including the merger of the WD with the core of the NS progenitor, and some types of bright transient events (intermediate luminosity optical transients; ILOTs). We note that in the case of a core-WD merger, the WD does not survive, and the system leaves only a NS. 

The idea that CCSN occurs inside a dense circumstellar matter (CSM) shell is not new (e.g. \citealt{Nelemansetal2001, Kalogeraetal2005, Churchetal2006, vanHaaftenetal2013}). However, in most (or even all) of these cases the shell originates from the exploding star itself or from a companion that did not form a WD yet. 
A PN is an expanding shell that the asymptotic giant branch (AGB) progenitor of the central star formed.
We note that less than one per cent of planetary nebulae are observed around a red giant branch star, (e.g., \citealt{Jonesetal2020R}). Therefore, for the purpose of our study we can refer to all planetary nebulae as being around post-AGB stars.  The central star of the planetary nebula is evolving to become a WD and it ionises the expanding shell. To have a CCSN inside a PN, which we term here CCSNIP, we need to have a binary system where the PN was formed before the CCSN, namely, a WD-NS reverse evolution where the WD forms before the NS (e.g., \citealt{SabachSoker2014}). More than that, we need the explosion of the secondary star to take place before the PN is dispersed in the interstellar medium (ISM). The dispersion of such a PN takes place at a distance of several parsecs, such that for a typical expansion velocity of $\approx 5-10 \km \s^{-1}$ the dispersion time is less than million years (e.g., \citealt{Napiwotzki2001, Wuetal2011}). Namely, for the formation of a CCSNIP we require the explosion to take place within about one million years (or even only $3 \times 10^5 \yr$) from the formation of the PN. 

The unique signature of a CCSNIP is the collision of the CCSN ejecta with a relatively dense shell in a stellar population where   a single star evolution will not lead to a CCSN nor to a dense CSM.  The question we examine is whether this might occur within an isolated binary system 
or whether a CCSNIP might occur only when the explosion of the reverse evolution occurs within the PN of another (a third) star in a triple-star system or in a young open cluster. To answer this question we use \textsc{MESA-binary} and \textsc{MESA-single} (Modules for Experiments in Stellar Astrophysics, section \ref{sec:simulations}) to simulate a grid of binary systems undergoing WD-NS reverse evolution, as we summarise in section \ref{sec:WhatWeDo}. 
We present our results in section \ref{sec:results}, and conclude with our main findings in section \ref{sec:summary}. 

\section{Numerical Setup} 
\label{sec:simulations}

We use \textsc{MESA~binary} (version 10398; \citealt{Paxtonetal2011, Paxtonetal2013, Paxtonetal2015, Paxtonetal2018, Paxtonetal2019}) to simulate the WD-NS reverse evolution. Conducting these simulations from start to end with \textsc{MESA~binary} is not straightforward (e.g., \citealt{GibsonStencel2018}). 
For example, \cite{GibsonStencel2018} use \textsc{MESA~binary} to explore the evolutionary state of the epsilon Aurigae binary system, where a F0 supergiant transfers mass to a companion. The nature of the companion is not clear. \cite{GibsonStencel2018} build a model with a binary system having initial masses of $9.85 M_\odot$ and $4.5 M_\odot$, and an orbital period of 100 days. Due to mass-transfer the present masses in their model are  $1.2 M_\odot$ and $10.6 M_\odot$, respectively. This large mass-transfer amount makes their study relevant to our goals. 

In about $60 \%$ of the simulations of \cite{GibsonStencel2018} the time step became too small or \textsc{MESA~binary} was unable to converge to an acceptable evolved stellar model.  These problems often occurred when the primary was on the AGB in their simulations. This large ratio of problematic simulations emphasises the difficulties in conducting such simulations.

To overcome the convergence problems we divide each run to two modes, the binary mode and the single star mode. 
Our subscript notation throughout this paper are as follows. The numbers 1 and 2 stand for the primary and secondary stars, respectively, ${\rm B}$ and ${\rm S}$ stand for the binary and single modes of \textsc{MESA}, respectively, and $ {\rm i}$, and ${\rm f}$ for the initial and termination points, respectively.
We also use the subscript $`{\rm MT}'$ to indicate a quantity at the beginning of the mass-transfer during the binary mode. 
In the binary mode, we use \textsc{MESA~binary} and terminate the run when the primary reaches a mass of $M_{\rm 1,B,f}=1.3 M_\odot$. At this point the primary is already on its route to transform to a WD, and we avoid numerical convergence problems that would appear later. We do not evolve the primary further.
We tried other limits such as $M_{\rm 1,B,f}=1.1 M_\odot$ and $M_{\rm 1,B,f}=1.2 M_\odot$ for numerous simulations. Some encountered problems (such as convergence and limitation in the time step) but most simulations converged. The final values of different variables in the binary mode were different due to the above change in the mass limit $M_{\rm 1,B,f}$ at the termination of the binary mode. However, the final outcome $\Delta t_{\rm CCSN}$ (see definition below) after the single mode of each simulation did not change much due to this change in $M_{\rm 1,B,f}$.

In the single star mode we use \textsc{MESA~single} and we simulate the secondary star that accreted mass from the primary star in the binary mode, and start the single mode with a mass of $M_{\rm 2,S,i} \ga 8 M_\odot$. We continue the single mode until the secondary star ends its evolution as a CCSN (leaving a NS), or as a WD if accretion takes place at late times (section \ref{sec:results}). Specifically, we terminate the secondary star evolution by one of four conditions, as follows. 
(1) $\log(T_c/K) >9.1$, where $T_c$ is the core temperature; (2) $\log(L_{\rm nuc}/L_\odot) > 10 $ where $\log(L_{\rm nuc})$ is the total power from all nuclear reactions; (3) $\log{L}< 0.1 L_\odot$; (4) $\log{R}< 1 R_\odot$. 
The first two conditions imply that there is oxygen burning in the core, and so the star is very close to core collapse and then NS formation. Conditions 3 and 4 indicate that the secondary star is deep in the post-AGB track on its way to form a WD (the second WD in the system). 

We hereby specify the general numerical details that hold for all simulations. In section \ref{sec:results} we list the unique properties for each run. 

We consider only circular orbits, and in the simulations we present in section \ref{sec:results} we do not take into account tidal forces. We performed 4 tests where we did include tidal forces, and compared with simulations having the same initial conditions but that did not include tidal forces. The initial masses and orbital periods for these four cases are ($M_{\rm 1,B,i}, M_{\rm 2,B,i}, P_i)=(7,6.98,200)$, $(6,5,100)$, $(6,5.6,100)$, and $(7,6.4,100)$, where masses are in $M_\odot$ and period is in days. 
The first three simulations showed very minor variations from the runs without tidal forces. The fourth run did not converge. The investigation of this issue in full is beyond the scope of this paper.
   
In the \textsc{MESA-binary} mode we follow the \textit{inlist} of \cite{GibsonStencel2018}, in simulating the stellar internal structure and binary evolution, like in taking their value for the rotation velocity, their mass-loss scheme, and in some cases their nuclear reaction net, and in particular we adopt the mass-transfer scheme of \cite{KolbRitter1990}. The mass-transfer efficiency scheme that we adopted is the one that \cite{GibsonStencel2018} use (from \citealt{Sobermanetal1997}), with the parameters: $\alpha=0.1$ for the fractional mass-loss from the vicinity of the donor star, lost as fast wind; $\beta=0.1$ for the fractional mass-loss from the vicinity of the accretor star, lost as fast wind; $\delta = 0.1$ for the fractional mass-loss from the circumbinary
coplanar toroid, with a radius equal to $\gamma^2 a$, where $a$ is the binary semi-major axis. We further follow \cite{GibsonStencel2018} and adopt $\gamma = 1.3$. 
 In summary, for each mass $\Delta M_1$ that the primary loses via a Roche lobe overflow (RLOF), a mass of $\alpha \Delta M_1=0.1\Delta M_1$ is lost from the primary vicinity, a fraction of $\beta \Delta M_1=0.1\Delta M_1$ is lost from the secondary star, and a fraction of $\delta \Delta M_1=0.1\Delta M_1$ is lost from a circumbinary torus. The mass that ends in the secondary star is $\eta \Delta M_1=0.7 \Delta M_1$, where $\eta=1-\alpha-\beta-\delta=0.7$. 

For the initial equatorial surface rotation velocity of both stars we take $v_{1,e}, v_{2,e} =2 \km \s^{-1}$. All other parameters except the nuclear reaction, stellar masses and orbital periods, are as in the \textit{inlist} of \cite{GibsonStencel2018}. For the nuclear reaction network we take the \textsc{MESA-binary} default;  we compared two successful runs with the nuclear reaction of \cite{GibsonStencel2018} to the default nuclear reaction of \textsc{MESA}. We found that the differences in the helium masses  of the primary stars at the end of the binary mode are about 1\% or less, and the differences in the helium masses of the secondary stars at the end of the single mode are 15\% and 3\% for the two cases we checked.   We also found that for all other parameters, in particular the time delay which is our main focus, the difference between the two cases was much less than a per cent. 

Other parameters that we do not mention here are set to their default option in \textsc{MESA-binary} or in \textsc{MESA-single}.  We note that the parameter space of the initial properties of the binary system, two masses, semi-major axis and eccentricity, and of the mass-transfer and mass-loss parameters, is huge. For example, if we change the mass of the primary star by $\Delta M= 0.5 M_\odot$ in the range of $M_{\rm 1,B,i}=6-8M_\odot$, the secondary mass by $\Delta M= 0.1 M_\odot$ in the range of $M_{\rm 2,B,i}=M_{\rm 1,B,i}-1 M_\odot$ to $M_{\rm 1,B,i}$, the orbital period by $\Delta P =50 \days$ from 50 to 250 days, and scan eccentricity values of of $e=0, 0.4,0.6$ and $0.8$, we have 1000 cases just of the initial conditions. We can then vary the fraction of mass lost by each star as RLOF takes place. In total, the parameter space is of thousands of cases to simulate. 

 Instead, we examine the volume of the parameter space that we think is the most favourable for forming a CCSNIP in a binary system. We cannot start with a primary star more massive than about $8.5 M_\odot$, as then it will end its life as a CCSN rather than a planetary nebula. As well, the changes from the cases with $M_{\rm 1,B,i}=6 M_\odot$ to the cases with $M_{\rm 1,B,i}=7 M_\odot$ make the CCSNIP less likely, as the time from planetary nebula formation to the CCSN of the secondary star does not decrease (section \ref{sec:results}).
Starting at a too close orbital separation will not help with the primary mass, as it will lead to a merger of the two stars before the primary evolves to form a planetary nebula.
We think that the values we examine for $M_{\rm 1,B,i}$ span the most favourable range to form CCSNIP.   

We also take favourable mass transfer parameters in the following way. For the secondary star to be massive enough to explode, and with the shortest time delay from the formation of the PN, we need it to accrete as much mass as possible from the primary. We therefore assume that the secondary star accretes most (a fraction of 0.7) of the mass that the primary star losses. It is not all the mass that the primary star losses because we do expect some mass to be lost, e.g., a wind from the primary star and mass loss to remove extra angular momentum from an accretion disk around the secondary star.We let only $10 \%$ of the mass to be lost in each case, i.e., $\alpha=0.1$, $\beta=0.1$, and $\delta = 0.1$, respectively. 

In the \textsc{MESA-single} mode, where we simulate the evolution of the secondary star from where it terminated the binary mode (in one case we also followed the primary star), we follow the parameters from the \textit{inlist} of \cite{GofmanSoker2019}. The initial structure (mass, radius, composition as function of radius) of the secondary star is taken directly from the last point of the binary mode, rather than starting at the main sequence or earlier.  
 
Before we discuss the results we point out the possibility that the system enters a common envelope evolution as the secondary star turns to a giant and swallows the WD remnant of the primary star.
We do not simulate this possibility here. \cite{SabachSoker2014} list 14 outcomes when a system performing the reverse evolution enters a common envelope evolution (CEE). Namely, when the secondary star engulfs the WD remnant of the primary star (see their evolutionary diagrams in their figures 1 and 2). In our quest to answer the question of whether a binary system performing the reverse evolution can form a CCSNIP, it does not matter whether the system enters a CEE or not. Firstly, the WD had already formed the PN (or not in some cases). 
Secondly, the WD remnant of the primary star will not change the destiny of the secondary core to form a WD or a NS even if it merges with the core of the secondary star
The reason for that is that the secondary star swallows the WD when it is long after the main sequence, and its helium core is well developed. Some channels might lead to other types of supernovae, i.e., thermonuclear run away of the WD or accretion induced collapse of the WD as it accretes mass from the core of the secondary star. These are speculative scenarios, and even if they take place, are very rare. In short, it is quite possible that the WD remnant of the primary star enters a CEE with the giant envelope of the secondary star. But if the secondary was on its way to form a CCSN, in the majority of the cases it will end its life as a CCSN, whether the WD survives the CEE or not.  

\section{Summary of the simulation scheme}
\label{sec:WhatWeDo}

 Before describing the results, we briefly summarise the evolution that we simulate (more details are in section \ref{sec:simulations}).  

(1) \textit{The binary mode.}  With \textsc{mesa binary} we follow the evolution that includes the following. (a) The evolution from the main sequence of both stars. (b) Mass-transfer between the two stars. (b) Mass-loss due to the mass-transfer, and in one test case mass-loss by winds. (c) In some cases tidal interaction that influences the orbital separation.

 We terminate the \textsc{mesa binary} when $M_{1,B,f}=1.3M_\odot$.

 (2) \textit{The single mode.} We leave the WD remnant of the primary star unchanged, and follow the evolution of the secondary star from the point it left the binary mode with \textsc{mesa-single}. This implies that we ignore orbital evolution and mass-transfer in the simulation in this mode. However, we will discuss the case when the WD remnant of the primary star enters the envelope of the secondary star at late times.

 (3) Registering the relevant outcomes.
Although this stage is not a simulation phase, it is very important to focus on the relevant parameters.
We require that (a) the primary forms a PN, (b) the secondary explodes as a CCSN, and  (c) the time from (a) to (b) is $\Delta_{\rm CCSN} \la 10^6 \yr$ to ensure that at CCSN explosion the PN shell is still around, even if at a very large distance. We will see that by far the most constraining parameter is $\Delta_{\rm CCSN}$. Practically, we will find that the binary systems can meet conditions (a) and (b), but not condition (c). 
   
\section{Results}
\label{sec:results}

\subsection{Relevant simulations}
\label{subsec:Simulations}
In Table \ref{tab:outcome} we list the properties of the simulations that we performed. We chose the parameters that we estimate are the most optimistic to form a CCSNIP in a binary system. 

\begin{table*}[]
\footnotesize, 
\centering
 \begin{tabular}{|m{0.75cm}|m{0.75cm}|m{0.55cm}|m{0.8cm}|m{1.1cm}|m{1.2cm}|m{1.0cm}|m{0.68cm}|m{0.8cm}|m{1.1cm}|m{1.45cm}|m{1.0cm}|m{1.1cm}|m{0.2cm}|}
\hline
1 & 2 & 3 & 4 & 5 & 6 & 7 & 8 & 9 & 10 & 11 & 12 & 13 & 14 \\ \hline
$M_{\rm 1,B,i}$ & $M_{\rm 2,B,i}$ & $P_{\rm B,i}$ & $M_{\rm 2,B,f}$ & $M_{\rm 1,B,f,H}$ & $M_{\rm 1,B,f,He}$ & $\Delta t_{\rm CCSN}$ & $\Delta t_{\rm PN}$ & $P_{\rm B,f}$ & $M_{\rm 2,S,f,He}$ & $M_{\rm 2, B,He[MT]}$ & $R_{\rm 2, B[MT]}$ & Outcome & C \\ \hline
 &  &  &  & Envelope & He core &  &  &  & He core & He core &  &  & \textbf{} \\ \hline
$M_\odot$ & $M_\odot$ & $\days$ & $M_\odot$ & $M_\odot$ & $M_\odot$ & $10^6\yr$ & $10^6\yr$ & $\days$ & $M_\odot$ & $M_\odot$ & $R_\odot$ &  & \textbf{} \\ \hline
6 & 5 & 100 & 8.29 & 0.14 & 1.15 & 13.46 &  & 524.13 & 2.41 & 0.0000 & 3.52 & WD-NS &  \\ \hline
6 & 5.2 & 100 & 8.49 & 0.20 & 1.09 & 9.69 &  & 555.42 & 1.92 & 0.0000 & 3.78 & WD-NS &  \\ \hline
6 & 5.4 & 200 & 8.69 & 0.04 & 1.26 & 18.69 &  & 1172.76 & 2.77 & 0.0000 & 4.98 & WD-NS &  \\ \hline
6 & 5.55 & 200 & 8.84 & 0.37 & 0.93 &  & 9.02 & 1220.29 & 1.15 & 0.6764 & 23.64 & WD-WD &  \\ \hline
6 & 5.6 & 100 & 8.89 & 0.13 & 1.17 &  & 4.46 & 618.64 & 1.20 & 0.0000 & 4.65 & (WD-NS/WD) & f \\ \hline
6 & 5.6 & 200 & 8.89 & 0.38 & 0.92 &  & 8.01 & 1236.70 & 1.13 & 0.8526 & 83.51 & WD-WD &  \\ \hline
6 & 5.65 & 200 & 8.94 & 0.04 & 1.26 &  & 7.06 & 1250.80 & 1.11 & 0.9515 & 68.66 & WD-WD &  \\ \hline
6 & 5.7 & 100 & 8.99 & 0.13 & 1.17 & 5.41 &  & 633.57 & 1.34 & 0.0000 & 4.96 & (WD-NS/WD) & f \\ \hline
6 & 5.7 & 200 & 8.99 & 0.38 & 0.92 &  & 6.18 & 1267.06 & 1.11 & 1.0399 & 74.85 & WD-WD &  \\ \hline
6 & 5.8 & 200 & 9.09 & 0.38 & 0.92 &  & 4.98 & 1298.26 & 1.08 & 1.1501 & 65.07 & WD-WD &  \\ \hline
6 & 5.9 & 100 & 9.19 & 0.13 & 1.17 & 3.12 &  & 665.18 & 1.30 & 0.0000 & 5.47 & (WD-NS/WD) & f \\ \hline
6 & 5.9 & 200 & 9.19 & 0.38 & 0.92 &  & 3.57 & 1329.49 & 1.04 & 1.2430 & 72.74 & WD-WD &  \\ \hline
6 & 5.95 & 200 & 9.24 & 0.38 & 0.92 &  & 1.51 & 1343.12 & 0.97 & 1.2388 & 76.28 & WD-WD &  \\ \hline \hline
7 & 6 & 100 & 9.99 & 0.00 & 1.30 & 11.21 &  & 737.89 & 3.23 & 0.0000 & 4.08 & WD-NS &  \\ \hline
7 & 6.2 & 100 & 10.19 & 0.00 & 1.30 & 9.43 &  & 775.20 & 3.43 & 0.0000 & 4.37 & WD-NS &  \\ \hline
7 & 6.4 & 100 & 10.39 & 0.00 & 1.30 & 5.74 &  & 812.14 & 3.13 & 0.0000 & 4.72 & WD-NS &  \\ \hline
7 & 6.6 & 150 & 10.59 & 0.34 & 0.96 & 7.81 &  & 1273.74 & 3.64 & 0.0000 & 5.18 & WD-NS &  \\ \hline
7 & 6.7 & 200 & 10.69 & 0.31 & 0.99 & 4.21 &  & 1734.99 & 1.24 & 1.2067 & 97.38 & (WD-NS/WD) & f \\ \hline
7 & 6.8 & 100 & 10.79 & 0.00 & 1.30 & 3.69 &  & 886.02 & 3.11 & 0.0000 & 5.68 & WD-NS &  \\ \hline
7 & 6.8 & 200 & 10.79 & 0.38 & 0.92 & 8.27 &  & 1777.20 & 3.60 & 0.0000 & 6.51 & WD-NS &  \\ \hline
7 & 6.9 & 100 & 10.89 & 0.00 & 1.30 & 2.85 &  & 904.13 & 2.99 & 0.0000 & 5.74 & WD-NS &  \\ \hline
7 & 6.9 & 200 & 10.89 & 0.31 & 0.99 &  & 2.53 & 1808.19 & 1.17 & 1.4083 & 97.94 & WD-WD &  \\ \hline
7 & 6.91 & 100 & 10.90 & 0.00 & 1.30 & 6.97 &  & 906 & 4.02 & 0.0000 & 5.70 & WD-NS &  \\ \hline
7 & 6.92 & 100 & 10.91 & 0.00 & 1.30 & 3.33 &  & 908 & 3.06 & 0.6739 & 5.62 & WD-NS/WD &  \\ \hline
7 & 6.925 & 100 & 10.92 & 0.00 & 1.30 & 2.94 &  & 908.79 & 2.92 & 0.6810 & 5.54 & WD-NS/WD &  \\ \hline
7 & 6.93 & 100 & 10.92 & 0.00 & 1.30 & 3.47 &  & 910 & 3.12 & 0.6881 & 5.41 & WD-NS/WD &  \\ \hline
7 & 6.95 & 200 & 10.94 & 0.32 & 0.98 &  & 0.73 & 1826.39 & 1.01 & 1.4528 & 102.64 & WD-WD &  \\ \hline
7 & 6.98 & 200 & 10.97 & 0.32 & 0.98 &  & 0.40 & 1837.35 & 1.00 & 1.4801 & 109.43 & WD-WD &  \\ \hline
\end{tabular}
\caption{The relevant simulations of this study. We devote section \ref{subsec:Simulations} to explain the meaning of all quantities. Some important quantities are as follows. The input parameters, the initial primary mass $M_{\rm 1,B,i}$, the initial secondary mass $M_{\rm 2,B,i}$, and the initial orbital period $P_{\rm B,i}$. The quantities $\Delta t_{\rm CCSN}$ and $\Delta t_{\rm PN}$ give the time from when the primary star forms its PN to the time when either the secondary star explodes as a CCSN and leaves a NS (`WD-NS' in the last column), or it forms a PN and leaves a WD remnant (`WD-WD' in the last column), respectively. }
\label{tab:outcome}
\end{table*}

Table \ref{tab:outcome} presents two sets of simulations, one with initial primary mass of $M_{\rm 1,B,i}=6 M_\odot$ and one with $M_{\rm 1,B,i}=7 M_\odot$. The second column lists the initial secondary masses. 
 
Column 3 presents the initial orbital period and is termed $P_{\rm B,i}$ (no primary or secondary notations are needed since this value is for the binary system).

Column 4 presents the final mass of the secondary star in the binary mode $M_{\rm 2,B,f}$, after it accreted mass from the primary star. By definition of our scheme, the initial secondary mass in the single mode is $M_{\rm 2,S,i}=M_{\rm 2,B,f}$. 

Columns 5 and 6 are the masses of the different primary components at the end of the binary mode, of the hydrogen-rich envelope $M_{\rm 1,B,f,H}$ and of the mass of the core including the helium layer $M_{\rm 1,B,f,He}$, respectively. The helium core boundary is taken at the outermost radius where the hydrogen mass fraction is $\leq 0.01$ and Helium mass fraction is $\geq 0.1$. As we terminate the binary evolution when the primary mass is $M_{\rm 1,B,f}=1.3M_\odot$, we have $M_{\rm 1,B,f,H}$ + $M_{\rm 1,B,f,He} = M_{\rm 1,B,f}=1.3M_\odot$ (up to the accuracy of the calculations and of rounding numbers).

Column 7 is relevant to cases when the secondary star explodes as a CCSN and leaves a NS remnant (as we indicate by `WD-NS' in the last column of the table). It presents the duration of the single mode $\Delta t_{\rm CCSN}=t_{\rm 2,S,f}-t_{\rm 2,B,f}$. Namely, the time from the termination of the binary mode, $t_{\rm B,f}$, i.e., when $M_{\rm 1,B,f}=1.3 M_\odot$ which is very close to the time the primary forms a WD, to the time when either the secondary star central temperature exceeds for the first time the value of $10^{9.1} \K$, or the nuclear core luminosity exceeds $10^{10} L_\odot$. These conditions are met just before it explodes as a CCSN and leaves a NS remnant (at that time the secondary has a massive oxygen core). 
 Specifically, the end time of the single mode $t_{\rm 2,S,f}$ in these cases is either when 
(1) $\log(T_c/K) >9.1$, where $T_c$ is the core temperature, or when (2) $\log(L_{\rm nuc}/L_\odot) > 10 $ where $\log(L_{\rm nuc})$ is the total power from all nuclear reactions, which ever occurs first.  

We note the following. 
For numerical reasons we stop the binary mode when $M_{\rm 1,B,f}=1.3 M_\odot$. By that time the primary star has a very massive core, so it is very luminous (see section \ref{subsec:HRdiagram}), implying a very high mass-loss rate of $\dot M_1 \approx {\rm few} \times 10^{-6} M_\odot$. Since the mass of the hydrogen-rich envelope is $M_{\rm 1,B,f,H} < 0.4 M_\odot$ (fifth column), the primary star will form a PN within $\approx 10^5 \yr$. This is a very short time relative to the other evolutionary times that we list in Table \ref{tab:outcome}, $\Delta t_{\rm CCSN}$ and $\Delta t_{\rm PN}$. The same consideration holds for the secondary star in cases where it forms a WD. Namely, the evolution time from when the secondary mass is $1.3 M_\odot$ to the formation of a PN is $\approx 10^5 \yr$. 
In some cases no hydrogen is left on the primary star when it is down to a mass of $M_{\rm 1,B,f}=1.3 M_\odot$. In such cases when the hydrogen is lost long before the central star ionises the nebula, the primary will not form a PN, or will form a faint PN if the hydrogen rich envelope is at large distances. This channel, of a WD but no regular PN, strengthens our main conclusion that binary systems without a tertiary star cannot form CCSNIPs.  

Column 8 refers only to cases where the secondary star leaves a WD remnant. It lists $\Delta t_{\rm PN}=t(M_{\rm 1,B,f}=1.3M_\odot)-t(M_{\rm 2,S,f}=1.3M_\odot)$, which is the duration of time between when $M_{\rm 1,B,f}=1.3 M_\odot$ in the binary mode (about the time when the primary forms a PN) and the time when the secondary mass reaches $M_{\rm 2,S,f}=1.3 M_\odot$ (about the time the secondary star forms a PN). 

Column 9 presents the orbital period of the binary system at the end of the binary mode $P_{\rm B,f}$. In all cases it is much longer than the initial orbital period, by a factor of $\approx 5-7$ for $M_{\rm 1,B,i} = 6 M_\odot$, and by a factor of $\approx 8-9$ for $M_{\rm 1,B,i} = 7 M_\odot$. This is a result of angular momentum balance (of the outflow and binary system together)  as the primary mass decreases by factor of about 4.6 and 5.4, respectively. Although the angular momentum of the binary system itself decreases because the mass that is lost from the system carries angular momentum, in the present study most of the mass that the primary losses ends on the secondary star.   

Column 10 represents the mass of the helium core of the secondary star at the end of the single mode, including the helium-rich layer, $M_{\rm 2,S,f,He}$. In cases when the secondary star leaves a WD remnant, this mass is very close to the final WD mass.

Column 11  and 12  list the mass of the helium core and the radius of the secondary star  when mass-transfer begins, $M_{\rm 2,B,He[MT]}$ and $R_{\rm 2,B[MT]}$, respectively.  From evolution time sequence considerations, columns 11 and 12 should to be between column 4 and 5. However, we place them before column thirteen to emphasise the connection of the core helium mass and secondary radius at mass-transfer to the fate of the secondary star, to be a WD or to be a NS, that we list in the column 13.  
In most cases when the secondary star has developed a massive helium core, $M_{\rm 2,B,He[MT]} \ga 0.7 M_\odot$, and a large radius, $R_{\rm 2,B[MT]} \ga 10 R_\odot$, namely it has left the main sequence at the onset of mass transfer, it will leave a WD remnant. In most cases when it is still on the main sequence when mass-loss begins, the secondary star will end in a CCSN and leave a a NS remnant.  
However, we also find that there are cases just on the boarder, where we cannot tell whether the secondary ends as a WD or as a CCSN that leaves a NS remnant. We mark these cases in column 13 by `WD-NS/WD'. The final outcome in these cases does not change the conclusions of this study, because the values of $\Delta t_{\rm CCSN}$ or $\Delta t_{\rm PN}$, whichever applies, are still large (see later). 

In Column 14 (`C'), we mark numerically-failed runs with an `f'. We can only estimate the fate of the failed runs, and so in column 13 we indicate our estimated fates inside parenthesis. We find, like in the successful runs, that also in the failed runs the time delay from PN formation to explosion is too long to form a CCSNIP in a binary system. 

We consider the maximum mass accretion rates of the two cases with $(M_{\rm 1,B,i},M_{\rm 2,B,i}) =(7 M_\odot, 6.8M_\odot)$. 
For $P_{\rm B,i}=200 \days$ the maximum accretion rate by the secondary star is $\dot M_{\rm 2,max} = 0.08 M_\odot \yr^{-1}$. During the time period when the mass accretion rate is $\dot M_{\rm 2,max} > 0.01 M_\odot \yr^{-1}$ the secondary mass grows from $M_2=7.1 M_\odot$ to $M_2=8.9 M_\odot$. For $P_{\rm B,i}=100 \days$ and the same initial masses, the maximum accretion rate is the same as for $P_{\rm B,i}=200 \days$, and during the time period when the mass accretion rate is $\dot M_{\rm 2,max} > 0.01 M_\odot \yr^{-1}$ the secondary mass grows from $M_2=7.0 M_\odot$ to $M_2=8.5 M_\odot$. We consider these accretion rates to be possible for massive main sequence stars. The reason is that we expect the mass-transfer to be through an accretion disk, and we expect the disk to launch jets (or blow a disk wind). The jets can remove energy and angular momentum and allow this high accretion rate. The possibility of this high accretion rate requires further studies.

During the binary mode of the simulations that we present in Table \ref{tab:outcome} we concentrated on mass-transfer, and therefore we turned off the stellar winds (we did include stellar winds in the single mode). To check the influence of the wind in the binary mode, we run the case of $(M_{\rm 1,B,i},M_{\rm 2,B,i},P_{\rm B,i}) =(7M_\odot, 6.8 M_\odot, 100 \days)$ with and without wind. We included the stellar winds as in  \cite{GofmanSoker2019}, and used the ”Dutch” scheme for massive stars wind mass-loss, with a mass-loss scaling factor of 0.8. The simulation with the stellar winds did not finish due to increased numerical run time. Nonetheless, we can conclude from what we managed to simulate that including stellar winds in the binary mode does not change much our results. Specifically, the inclusion of stellar winds changes the values of the final secondary mass in the binary mode, $M_{\rm 2,B,f}$, and the time delay $\Delta t_{\rm CCSN}$, by less that $1\%$ from the run without stellar winds. 

We examined the role of the final primary mass in the binary mode. For all runs in Table \ref{tab:outcome} this value is $M_{\rm 1,B,f} = 1.3M_\odot$. 
We also run cases with $M_{\rm 1,B,f}=1.2M_\odot$ and $1.1M_\odot$ for the initial parameters  $(M_{\rm 1,B,i},M_{\rm 2,B,i},P_{\rm B,i})=(6, 5, 100)$, $(6, 5.7, 200)$,  $(6, 5.9, 200)$, and $(7 M_\odot, 6.98 M_\odot, 200 \days)$. The values $M_{\rm 2,B,f}$ and $\Delta t_{\rm CCSN}$ in these runs differ from the runs with $M_{\rm 1,B,f}=1.3M_\odot$ by less than $1\%$. We performed one simulation with an initial orbital period of only $P_{\rm B,i}=70\days$, $(M_{\rm 1,B,i},M_{\rm 2,B,i},P_{\rm B,i})=(7M_\odot, 6.8M_\odot, 70 \days)$.  We did not finish this run because of tremendously increased simulation time, and had to stop the simulation in the \textsc{MESA-binary} mode when $M_{1,B}=1.334M_\odot$.  At the point where we terminated this run the values of $M_{\rm 2,B,f}$ and $\Delta t_{\rm CCSN}$ differ by less than $1 \%$ from the run with the same initial masses and an initial orbital period of $P_{\rm B,i}=100 \days$.

\subsection{The lessons from the simulations}
\label{subsec:Main}

The main points to take from Table \ref{tab:outcome} are as follows. 

(1) For the several cases with orbital periods of $P_{\rm B,i}=100-200 \days$ that we study we learn that if mass-transfer takes place after the secondary star has developed a massive helium core, $M_{\rm 2,B,He[MT]}$, the secondary star leaves a WD remnant, namely the binary system ends as a wide WD-WD binary system. On the other hand, when neglecting further mass-transfer events if the secondary did not develop a helium core by the time mass-transfer is initiated, it explodes as a CCSN (as long as its mass after mass-transfer is $M_{\rm 2,B,f} \ga 8 M_\odot$, which holds for all the relevant cases here). 

For the cases of initial primary masses and orbital separations we study here, the secondary star leaves a WD remnant for $5.5 M_\odot \la M_{\rm 2,B,i} < M_{\rm 1,B,i}=6 M_\odot$  and  $6.95 M_\odot \la M_{\rm 2,B,i} < M_{\rm 1,B,i}=7 M_\odot$, respectively. 

(2) The time period $\Delta t_{\rm CCSN}$ that we list in column 7 of Table \ref{tab:outcome} is the time from the formation of a PN by the primary star to the time when the secondary star explodes as a CCSN and leaves a NS remnant (column 13th).
We see that in all cases this time period is longer than the expected maximum time a PN might preserve its identity, which is at most $\approx 10^6 \yr$ (e.g. \citealt{Napiwotzki2001, Wuetal2011}).
The conclusion is that the reverse evolution cannot lead a binary system to explode a CCSN inside a PN. Namely, a reverse evolution of a binary system cannot form a CCSNIP. 

 This conclusion is an extrapolation from the volume of the parameter space of the binary stellar properties (initial masses, initial orbital separation, fraction of mass that is lost from the system, etc.) that we do cover with our simulations, to the volume of the parameter space that we do not cover. We justify this because we chose to simulate cases with the most optimistic values to obtain the lowest possible values of $\Delta t_{\rm CCSN}$.  (a) More massive primary stars than we use here will not form a WD, but rather explode as CCSNe. (b) Lower mass primary stars imply lower post-mass-accretion secondary star that do not lead to CCSN progenitors. (c) Lower mass secondary stars than we use here for a specific primary star,  will take even longer to form CCSNe, increasing the values of $\Delta t_{\rm CCSN}$. (d) A larger fraction of mass-loss out of the mass that the primary loses will not bring the post-mass-transfer secondary star to be a CCSN progenitor. (e) Larger initial orbital separations will not result in high enough mass-transfer to bring the post-mass-accretion secondary stars to be CCSN progenitors. (f) A much too short orbital separation will bring the two stars to merge to one star before the formation of a planetary nebula.

In section \ref{sec:summary} we discuss the way to form CCSNIP with the presence of a third star.    
   
(3) The hydrogen mass at explosion (not shown in Table \ref{tab:outcome}) in the different cases where the secondary star  explodes as a CCSN is in the range of $M_{\rm 2,H,exp}=3.7-5 M_\odot$ (the hydrogen-rich envelope mass is about 1.5 times larger).  However, we did not include the binary evolution in the single mode. In cases where the WD remnant of the primary star does not influence much the mass-loss from the secondary star, it will end as a type II CCSN. In most of these cases (but not all, due to a possible natal kick in the right magnitude and direction) the binary system, now of a WD and a NS, becomes unbound. This is since the mass that is lost at explosion, $M_{\rm ejecta} > 5.4 M_\odot$, is larger than the total mass of the WD-NS remnants. 
As said, the WD companion might enhance the mass-loss from the secondary star, by tidal spinning-up its envelope, by RLOF, or by a CEE. In that case the outcome will be a stripped-envelope CCSN (type IIb, type Ib, or even type Ic). The exact nature of the CCSNe is not the main focus of our study.

(4) The time period $\Delta t_{\rm PN}$ that we list in column 8 of Table \ref{tab:outcome} is the time from the formation of a PN by the primary star to the formation of a PN by the secondary star when it leaves a WD remnant (13th column). The values of $\Delta t_{\rm PN}$ for the cases $( M_{1,B,i},  M_{2,B,i}) = (6, 5.95)$ and $(M_{1,B,i}, M_{2,B,i}) = (7, 6.98)$ show that a fine tuning  (very close initial masses between the two stars)  might lead to a PN that the secondary star forms inside the very old PN that the primary star had formed. 
In these two cases when the primary reaches the end of the AGB it has sufficient hydrogen-rich envelope, $M_{\rm 1,B,f,H}=0.38M_\odot$ and $0.32 M_\odot$, respectively, to form a PN.  In our study we have ignored possible binary interactions after the primary has become a WD, but if this would be taken into account, then it will not prevent the formation of the second PN . Actually we think that most PNe are formed by binary interaction (e.g., \citealt{DeMarcoIzzard2017PASA} and references therein).

\subsection{Evolution on the HR diagram}
\label{subsec:HRdiagram}

\begin{figure*} 
\vskip -5.00 cm
\hskip -2.00 cm
\includegraphics[width=1.2\linewidth]{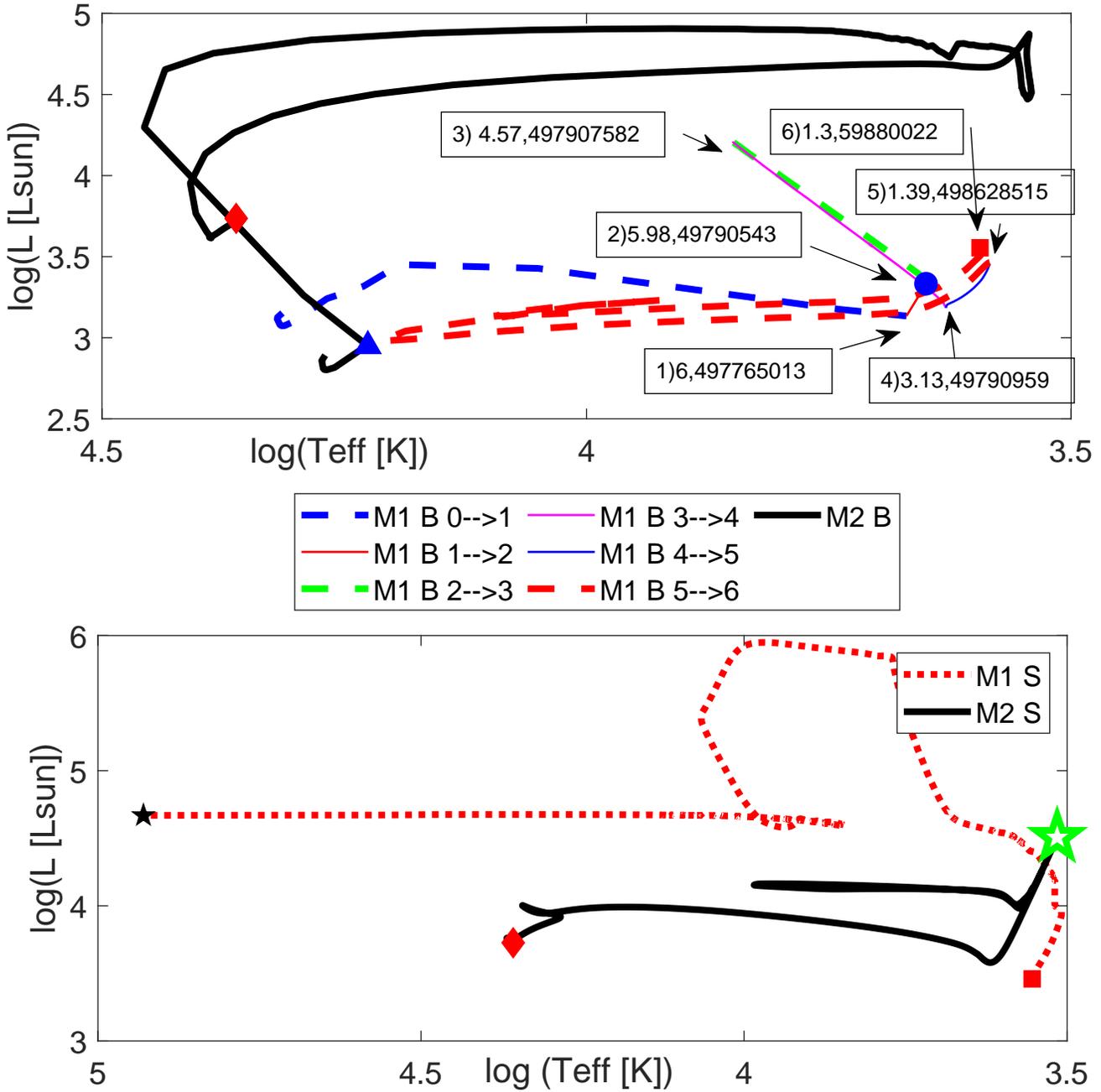}
\vskip -4.50 cm
\caption{The HR diagram for the system with $M_{\rm 1,B,i }=6M_\odot$, $M_{\rm 2,B,i}=5M_\odot$, and $P_{\rm i}=100\days$. The segmented line (each segment is listed in the figure itself) and the solid black line present the evolution of the primary and of the secondary stars, respectively in the upper panel. In the lower panel, the dotted red line and the solid black line represent the evolution of the primary and of the secondary stars, respectively
\newline
The top panel presents the HR diagram of the binary mode from \textsc{MESA~binary}, up to the formation of a primary star of mass $M_{\rm 1,B,f}=1.3M_\odot$ that occurs at $t_{\rm B,f}=6 \times 10^7 \yr$. 
We mark four points on the upper panel. 
(1) A red square at $t_{\rm B,f}$ marks the termination point of the primary evolution in the binary mode, with a mass of $M_{\rm 1,B, f}=1.3M_\odot$.
(2) A red diamond at $t_{\rm B,f}$ marks the termination point of the secondary evolution in the binary mode, with a mass of $M_{\rm 2,B,f}=8.3 M_\odot$ (see also Table \ref{tab:outcome}). 
(3) A blue circle at $t_{\rm MT}=5 \times 10^7 \yr$ marks the location of the primary star on the HR diagram when mass-transfer starts. (4) A blue triangle at $t_{\rm MT}$ marks the location of the secondary star on the HR diagram when mass-transfer starts. 
We mark with arrows six points on the evolution track of the primary star, and give the mass in $M_\odot$ and time in years. The evolution of the primary in the binary mode is differentiated to 7 segments according to these points as noted by the legends.  
\newline
The bottom panel presents the HR diagram of the single mode for the primary star and for the secondary star. The starting points are the final points in the upper panel (a red square and a red diamond, respectively). We arbitrarily set the termination point of the primary star evolution when its radius is $R_1=1 R_\odot$ (black pentagram), about the time it strongly ionises its PN. We end the evolution of the secondary star when $\log(L_{\rm 2, nuc}/L_\odot) = 10 $ (section \ref{sec:simulations}), very close to its explosion as type II CCSN (large open green pentagram). }
\label{fig:6_5_100}
\end{figure*}

\begin{figure*},
\vskip -6.00 cm
\hskip -2.00 cm
\includegraphics[width=1.2\linewidth]{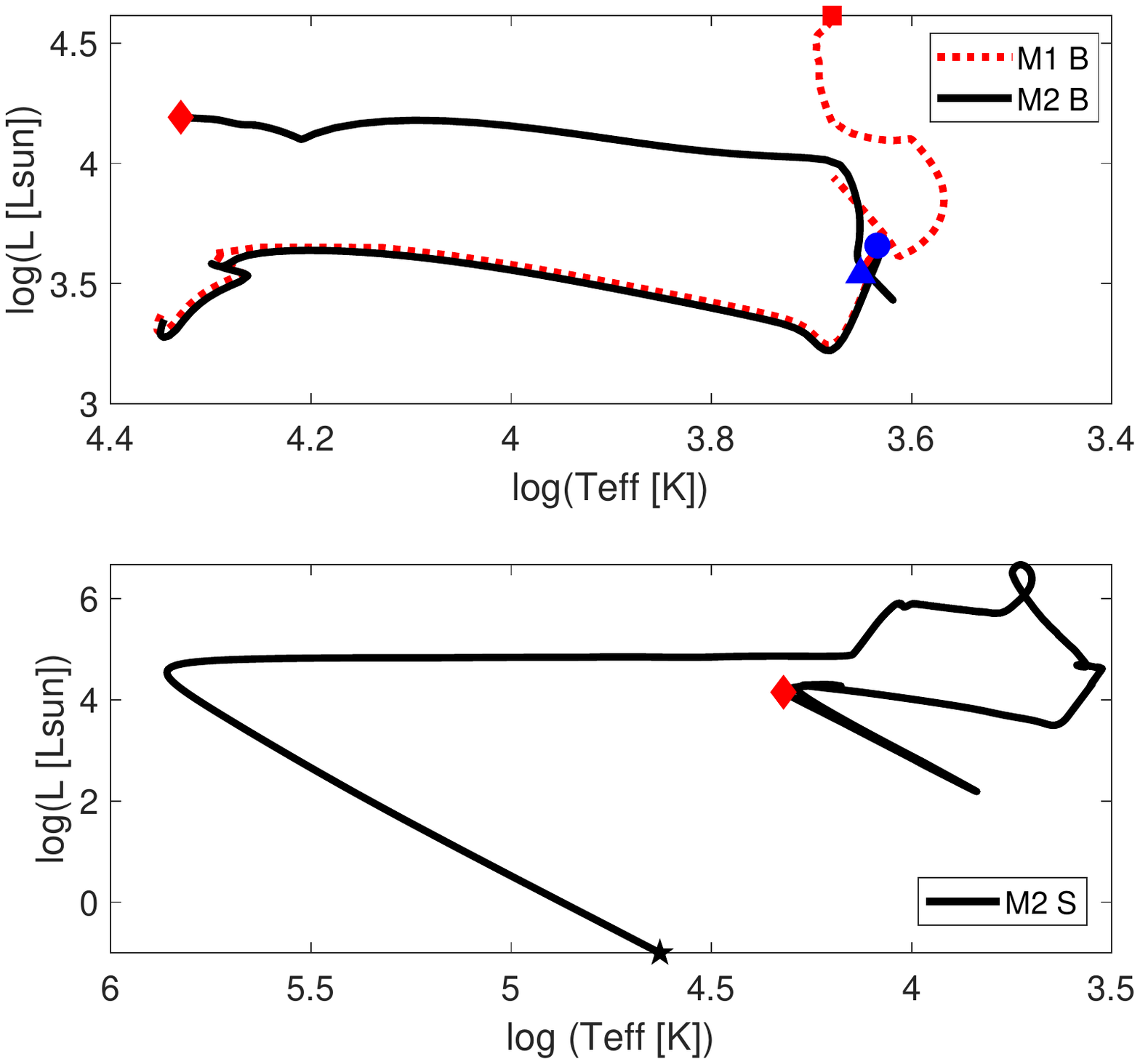}
\vskip -5.00 cm
\caption{The HR diagram for the binary system of $M_{\rm 1,B,i }=7M_\odot$, $M_{\rm 2,B,i}=6.9M_\odot$ and $P_{\rm i}=200\days$, which ends as a WD-WD wide binary system. The dotted red line and the solid black line present the evolution of the primary and of the secondary stars, respectively. 
\newline
The top panel presents the HR diagram of the binary mode from \textsc{MESA~binary}, up to the formation of a primary star of mass $M_{\rm 1,B,f}=1.3M_\odot$ that occurs at $t_{\rm B,f}=4.3 \times 10^7 \yr$. 
We mark four points on the upper panel. 
(1) A red square at $t_{\rm B,f}$ marks the termination point of the primary evolution in the binary mode, with a mass of $M_{\rm 1,B,f}=1.3M_\odot$.
(2) A red diamond at $t_{\rm B,f}$ marks the termination point of the secondary evolution in the binary mode, with a mass of $M_{\rm 2,B,f}=10.9 M_\odot$ (see also Table \ref{tab:outcome}). 
(3) A blue circle at $t_{\rm MT}=4.19 \times 10^7 \yr$ marks the location of the primary star on the HR diagram when mass-transfer starts. (4) A blue triangle at $t_{\rm MT}$ marks the location of the secondary star on the HR diagram when mass-transfer starts. 
\newline
The bottom panel presents the HR diagram of the secondary star in the single mode from $t_{\rm 2,S,i}=t_{\rm B,f}$, when the secondary mass is $M_{\rm 2,S,i} = M_{\rm 2,B,f}=10.9 M_\odot$ (red diamond), to the time when the secondary luminosity is $L_{2} = 0.1 L_\odot$, namely, when it is already on its WD cooling track (black pentagram). 
The secondary reaches the black pentagram location at $7.4 \times 10^7 \yr$ from the beginning of the binary mode (the zero age main sequence of the two stars).   }
\label{fig:7_6.9_200}
\end{figure*}

We present the evolution on the HR diagram for two systems, one that ends with a CCSN that leaves unbound WD and NS remnant (Fig. \ref{fig:6_5_100}), and one that leaves a wide WD-WD binary system (Fig. \ref{fig:7_6.9_200}). The primary and secondary stars have complicated tracks on the HR diagram that include some loops. As our aim in this study is to explore the final outcome, we do not analyse in detail that evolution (there are other studies of mass-transfer in different kinds of stars in the literature, e.g., \citealt{Poelarendsetal2017, Yoonetal2017, GibsonStencel2018, Wuetal2018, Brinkmanetal2019, Farrelletal2019, Gosnelletal2019}). 

The relevant properties to our study that Figs. \ref{fig:6_5_100} and \ref{fig:7_6.9_200} reveal are as follows. 

(1) In our simulations there is a mass-transfer episode from which most of the mass is accreted by the secondary star. This mass-transfer episode that later leads the secondary star to  explode as a CCSN takes place while the secondary star is still on the main sequence, and did not develop a helium core yet (blue triangle on the upper panel of Fig. \ref{fig:6_5_100}).  
During and shortly after the mass-transfer process the secondary performs a large loop on the HR diagram (black line on the upper panel of Fig. \ref{fig:6_5_100}), and returns to the main sequence as a more massive star (red diamond). 
It later evolves towards a CCSN (black line on the lower panel of Fig. \ref{fig:6_5_100}). 

(2) In the cases  that we simulated  where the secondary star explodes as a CCSN, it is a red supergiant at explosion (large open green pentagram on the lower panel of Fig. \ref{fig:6_5_100}).  However, as we indicated above, we did not consider cases where the WD remnant of the primary star enhances mass-loss from the secondary star to bring it to explode as a stripped-envelope CCSN. In these later cases the exploding star might be a blue star. 

(3) If mass-transfer takes place after the secondary star has left the main sequence (blue triangle on the upper panel of Fig. \ref{fig:7_6.9_200}), it ends as a WD. It contracts after accreting mass, and then resumes evolution to become an AGB star and to form a PN (lower panel of Fig. \ref{fig:7_6.9_200}).

\section{Discussion and Summary}
\label{sec:summary}

\subsection{General outcomes}

The question we raised at the beginning of this study was whether a single binary system that experiences the WD-NS reverse evolution, where the WD forms before the NS, might form a CCSN inside a PN, so called CCSNIP. 
  
To perform a reverse evolution, the primary star of the system should have a zero age main sequence mass of $M_{\rm 1,B,i} \la 8-9 M_\odot$, such that it forms a PN and leaves a WD remnant. When the primary star evolves it transfers mass to the secondary star, such that after this binary interaction with mass-transfer the secondary mass becomes $M_{\rm 2,B,f} \ga 8 M_\odot$. To explode as a CCSN, it is also necessary that the secondary star mass grows to $M_{\rm 2,B,f} \ga 8 M_\odot$ before it had developed a helium core (Table \ref{tab:outcome}). We present one such case in Fig. \ref{fig:6_5_100}. 
  
In cases where the mass-transfer takes place after the secondary star has developed a helium core ($M_{\rm 2,B,He[MT]} > 0$ in Table \ref{tab:outcome}) and left the main sequence, it forms a PN rather than a CCSN, and leaves a WD remnant, as we show for one case in Fig. \ref{fig:7_6.9_200}.  If the WD remnant of the primary star does not enter the envelope of the secondary star as the secondary becomes a giant,  the remnant of such a binary system is a bound wide WD-WD binary system.  If the WD does enter the envelope of the secondary star to form a CEE, then there are two possible outcomes. In the first outcome the WD survives the CEE and the core of the secondary star forms a second WD. The final orbital separation of this post-CEE binary system is much smaller than the initial one and can be as short as  $\approx R_\odot$. In the second outcome of the CEE the WD merges with the core to form a massive WD merger-product. This might be a progenitor of a SNe Ia in the frame of the core degenerate scenario \citep{KashiSoker2011}, might experience a collapse to form a NS, or might form a rapidly rotating magnetic WD (e.g., \citealt{Wickramasingheetal2014}).  

When the initial masses of the two stars are very close, the time period from the formation of the first PN by the primary star to the formation of the second PN by the secondary star is $\Delta t_{\rm PN} < 10^6 \yr$ (last two rows of Table \ref{tab:outcome}). Since some PNe can retain their identity for hundreds of thousands of years, it is possible that in such cases we form a PN inside a large and old PN. \cite{Lopezetal2000} already suggested that KJPN~8 is composed of two consecutive PNe that originated from a binary system that had very similar initial masses. In the case of KJPN~8, however, the time period between the formation of the two PNe is only $\approx 10^4 \yr$, which requires a hyper-fine-tuned set of initial masses.  

In cases of reverse evolution  that we studied  we found that the secondary star explodes when it is a red supergiant, and it forms a SN II. The explosion unbinds the WD and the NS remnants,  unless in rare cases natal kick keeps the binary bound. 
 In addition, in cases that we did not simulate here where the WD remnant of the primary star enhances mass-loss from the secondary star, the secondary star might explode as a more compact bluer stripped-envelope CCSN. 
 
Most relevant to our study is our finding that we could not bring the time period from the formation of the PN by the primary star to the explosion of the secondary star, $\Delta t_{\rm CCSN}$, to be less than a few millions years (column 7 of Table \ref{tab:outcome}). If we set the two initial masses to be too close to each other, the secondary leaves the main sequence before mass-transfer takes place, and it does not explode as a CCSN. Our finding that $\Delta t_{\rm CCSN} > 10^6 \yr$ implies that by the time the secondary explodes as a CCSN, the PN had long dispersed into the ISM (section \ref{subsec:Main}).  

We highlight two interesting outcomes for the systems we have considered so far. The first involves a thermonuclear runway, like in SNe Ia. This might take place when the CO WD remnant of the primary star accretes mass and explodes as a SN Ia, or, more likely, that it merges with the core and the two suffer a SN Ia \citep{SabachSoker2014}. In both cases, if this explosion takes place inside the H-rich envelope (and core), the explosion mimic a CCSN. This is similar to the scenario that leads to a SNIP, i.e., a SN Ia inside a planetary nebula (the regular SNIP scenario does not require a reverse evolution, and for that can take place in lower mass stars).  
The second channel involves AIC, where the ONeMg WD remnant of the primary star accretes mass from the secondary core and collapses to form a NS inside the core (or envelope) of the secondary star when the secondary is a giant \citep{SabachSoker2014}.  Namely, the core and the ONeMg WD merge.  Such an AIC releases a large amount of gravitational energy that can eject most (or all) of the core and envelope and mimic a CCSN. This leads to an explosion that leaves a NS remnant and occurs after a PN formation. However, as with the other scenarios we study here, we expect that the delay time will be too long between PN formation and the AIC-explosion.

We suggest that the formation of a CCSNIP requires a third star in the system. 
The third star can be a very wide tertiary star in the system (such that it evolves independently) but at an orbital separation of $a_3 \la 1 \pc$, or can be a member in an open stellar cluster that is not too far from the binary system (again, a distance of $D_3 \la 1 \pc$).  
The binary system performs the WD-NS reverse evolution and leads the secondary star of initial mass of $M_{\rm 2,B,i} < 8 M_\odot$ to explode as a CCSN. The third star has a mass that is in between the initial masses, $M_{\rm 2,B,i} < M_{\rm 3,i} < M_{\rm 1,B,i}$, and it forms a PN within few hundreds of thousands of years before the secondary star explodes as a CCSN, $\Delta t_{\rm PN3,CCSN} \la 5 \times 10^5 \yr$. The PN might preserve its identity, although deformed by the ISM, by the time the secondary star explodes. In case the third star is a cluster member the CCSN might take place far from the center of the PN. 

\subsection{ The fraction of CCSNIP}

We crudely estimate the fraction of CCSNIP relative to all CCSNe as follows. 
\cite{SabachSoker2014} estimated that the event rate of all routes of WD-NS reverse evolution is $f_{\rm RE} \approx 3-5 \%$ of the CCSN rate.
 \cite{SabachSoker2014} calculated this rate from an initial mass function and taking the initial primary mass to be in the range of $5.5 M_\odot < M_{\rm 1,B,i} < 8 M_\odot$. They also assumed a flat mass ratio distribution in the binary systems and that $60 \%$ of the primary stars are in binaries. They then estimated that the probability of a binary system to be
in the desired orbital separation is $20 \%$. Integrating over the allowed parameter space, they derived their estimate of $f_{\rm RE} \approx 3-5 \%$  (they did not perform population synthesis).  
Note that  it is irrelevant to our research question  whether the WD survives the evolution (as the cases we studied here), or whether in enters a common envelope evolution with the secondary star as the later becomes a giant (\citealt{SabachSoker2014, Soker2019SCPMA}).  
From \cite{MoeDiStefano2017} we find that for the initial mass range here of $M_{\rm 1,B,i} \simeq 5.5-8 M_\odot$ the fraction of single, binary, triple, and quadruple stars are about 0.24, 0.36, 0.27, and 0.13, respectively. This implies that on average each binary system has $f_{\rm 3,B,t} \simeq (0.27+2\times 0.13)/(0.36+0.27+0.13)=0.7$ extra stars in triple (or quadruple) bound stellar system. It is more difficult to estimate the presence of an extra star from the open cluster, $f_{3,B,c}$. We simply assume that the cluster members add somewhat to this fraction that becomes larger than $f_{\rm 3,B,t} \simeq 0.7$. We therefore crudely assume that on average each binary system has about one extra star that serves as the third star in the system that forms the PN shortly before the secondary star explodes as a CCSN, $f_{3,B} \approx 1$. Namely, the cluster members contribute only about $30 \%$,  $f_{3,B,c} \simeq 0.3$, of the extra stars (as multiple systems contribute 0.7 extra stars for each binary system). 

From the values of $\Delta t_{\rm PN}$ in Table \ref{tab:outcome} and from the duration of the evolution of stars until they form a WD as function of their initial mass (e.g., \citealt{Ekstrometal2012, Choietal2016}), we find that the third star (tertiary or an open cluster member) should have an initial mass within $\Delta M_{2,3} \approx 0.015-0.04 M_\odot$ of the secondary star to form a PN within about $7 \times 10^5 \yr$ before the secondary star explodes.
The value of $\Delta M_{2,3} \approx 0.015 M_\odot$ is for $M_{\rm 2,B,i} = 5 M_\odot$, and that of $\Delta M_{2,3} \approx 0.04 M_\odot$ is for $M_{\rm 2,B,i} = 7 M_\odot$. Namely, $M_{\rm 2,B,i}-\Delta M_{2,3} < M_{\rm 3,i} < M_{\rm 2,B,i}$. For a flat mass distribution of the tertiary star, the probability for this mass range is $f_{3,M} \approx \Delta M_{2,3}/M_{\rm 2,B,i} \approx (0.015M_\odot/5M_\odot) - (0.04M_\odot/7M_\odot) \simeq 0.003-0.006$.

Overall, we crudely estimate the fraction of CCSNIP events from all CCSNe to be 
$f_{\rm CCSNIP} \approx f_{\rm RE} f_{3,B} f_{3,M} \approx 10^{-4} - 10^{-3.5}$. 
As future surveys aim at about $10^4$ CCSNe per year or so, we expect that about one to few of these will be CCSNIP, i.e., CCSN inside an old PN. The interaction of the ejecta with the relatively dense PN might take place tens to hundreds of years after explosion. The density of the PN is expected to be larger than that expected for CCSN in old open clusters.   
A more accurate estimate of the event rate of CCSNIPs and their possible observational signatures are the subjects of future studies.

\acknowledgments
We thank an anonymous referee for detailed comments that improved the manuscript.  This research was supported by a grant from the Israel Science Foundation. We completed this work while the Technion was closed due to the Coronavirus (COVID-19).

\textbf{Data availability}
The data underlying this article will be shared on reasonable request to the corresponding author.  

\pagebreak

\end{document}